\documentstyle[12pt]{article}
\begin{document}

\begin{titlepage}

\hfill{August 1995}

\hfill{}

\hfill{UM-P-95/75}

\hfill{RCHEP-95/19}

\vskip 1 cm

\centerline{{\large \bf
Reconciling sterile neutrinos with big bang nucleosynthesis
}}

\vskip 1.5 cm

\centerline{R. Foot and R. R. Volkas}

\vskip 1.0 cm
\noindent
\centerline{{\it Research Centre for High Energy Physics,}}
\centerline{{\it School of Physics, University of Melbourne,}}
\centerline{{\it Parkville, 3052 Australia. }}

\vskip 1.0cm

\centerline{Abstract}
\vskip 1cm
\noindent
We re-examine the big bang nucleosynthesis (BBN) bounds on
the mixing of neutrinos with sterile species.
These bounds depend on the assumption that the
relic neutrino asymmetry $L_{\nu}$ is very small.
We show that for $L_{\nu}$ large enough (greater than
about $10^{-5}$) the standard BBN bounds do not apply.
We apply this result to the sterile neutrino
solution to the atmospheric neutrino anomaly and show that
for $L_{\nu} > 7 \times 10^{-5}$ it is consistent with BBN.
The BBN bounds on sterile neutrinos mixing with electron
neutrinos can also be weakened considerably.
\end{titlepage}

The solar neutrino deficit \cite {snd},
atmospheric neutrino anomaly \cite{ana},
and LSND experiment\cite{lsnd} can all be viewed as
evidence for
non-zero neutrino masses and oscillations.
It does not seem possible to explain all these anomalies
with the three known neutrino species and thus
new neutrino species might exist.
Given that new ordinary weakly interacting neutrino species
are ruled out by LEP, sterile neutrinos ($\nu_s$) are a natural
candidate.  There are essentially two types of sterile
neutrinos that can be envisaged.
Firstly, there are sterile states which either
have no gauge interactions, or interactions which are
much weaker than the usual weak interactions \cite{sn}.
Alternatively, it is possible to envisage neutrinos which
do not have significant interactions with ordinary matter
but do have significant interactions with themselves.
An interesting example of the latter is
given by mirror neutrinos which interact with themselves only
through mirror weak interactions which have the same strength as
ordinary weak interactions \cite{flv}.

However, for both sterile and mirror neutrinos
there are apparently quite stringent bounds
if they are required to be
consistent with standard big bang cosmology.
Assuming that the number of effective
neutrino species present during nucleosynthesis is bounded to be
less than 4, then the mixing angle ($\theta_0$) and the squared
mass difference ($\delta m^2$) for a sterile neutrino mixing with
one of the known neutrinos is bounded by
(assuming $\delta m^2 > 0$)\cite{B}:
$$\delta m^2 \sin^4 2\theta_0\
\stackrel{<}{\sim}
\ 5 \times
10^{-6} \  eV^2 ,\   \nu = \nu_e,  $$
$$ \delta m^2 \sin^4 2\theta_0\ \stackrel{<}{ \sim}\  3\times
10^{-6} \ eV^2, \  \nu=\nu_{\mu, \tau}. \eqno (1)$$
These bounds arise by demanding that oscillations
do not bring the sterile neutrino into equilibrium with the
known neutrinos. Electron neutrinos must also not be depleted
too much by oscillations after decoupling,
 during the BBN epoch,
because then the freeze out temperature for neutron-proton transitions
is increased. For maximal
mixing, the bound on $\delta m^2$ is extended to
$\delta m^2\ \stackrel{<}{\sim}\ 10^{-8}\ eV^2$ \cite{B}.
These ``bounds'' would appear to exclude the region of parameter
space required to explain the atmospheric neutrino anomaly
in terms of $\nu_{\mu}-\nu_s$ oscillation ($\delta m^2 \simeq 10^{-2}
eV^2$, $\sin^2 2\theta_0 \simeq 1$),
and would restrict the parameter space required to explain
the solar neutrino deficit in terms of $\nu_e - \nu_s$ oscillation.
An important assumption in deriving the bounds Eq.(1) is that
the relic neutrino asymmetries could be neglected. However, the
neutrino asymmetries cannot be measured, and at present
the origin of particle asymmetries is not fully understood.
Only the extremely weak bound $L_{\nu_{\alpha}}\ \stackrel{<}{\sim}\
10^3$ can be derived by demanding that the
neutrinos do not violate the upper
limit on the total energy density of the universe.
The purpose of this letter is to re-examine the BBN bounds
on ordinary-sterile neutrino mass and mixing for
arbitary neutrino asymmetries.
In particular, we will show that for neutrino asymmetries
larger than about $7 \times 10^{-5}$
the standard big bang model is consistent with
sterile neutrinos mixing with muon neutrinos with parameters
suggested by the atmospheric neutrino anomaly \cite{babu}.

Let us first examine the ordinary neutrino ($\nu_{\alpha}$,
$\alpha = e, \mu, \tau$) oscillating
with a sterile neutrino ($\nu_s$) in vacuum. Oscillations can occur
if the weak eigenstate
neutrino and sterile neutrino are each
linear combinations
$\nu_{\alpha} = \cos\theta_0 \nu_1 + \sin\theta_0 \nu_2 $ and
$\nu_s = -\sin\theta_0 \nu_1 + \cos \theta_0 \nu_2$ of mass
eigenstates $\nu_{1,2}$.
An ordinary neutrino of momentum $p$
will then oscillate in vacuum after a time $t$
with probability
$$|\langle \nu_{\alpha} (t)|\nu_s\rangle|^2 = \sin^2 2\theta_0
\sin^2 \left({t \over L_{osc}}\right), \eqno (2)$$
where \cite{c2}
$$L_{osc} = {2p \over \delta m^2} \equiv {1 \over \Delta_0}.\eqno (3)$$
However, in the early universe oscillations occur in a plasma.
For $\nu_{\alpha} - \nu_s$ ($\alpha = e, \mu, \tau$)
oscillations in a plasma of temperature $T$, the matter and vacuum
oscillation parameters are related by \cite{msw}
$$\sin^2 2\theta_m = {\sin^2 2\theta_0 \over 1 - 2z \cos 2\theta_0
+ z^2},$$
$$\Delta_m^2 = \Delta_0^2 ( 1 - 2 z \cos 2\theta_0 + z^2 ),
\eqno (4)$$
where $z = 2\langle p \rangle
\langle V_{\alpha} - V_s \rangle /\delta m^2$
and $\langle V_{\alpha, s} \rangle$ are the effective potentials
due to the
interactions of the neutrinos with matter ($\langle p \rangle \simeq 3.15
T$). For a truly sterile
neutrino $V_s = 0$. (For neutrinos which have only self
interactions, e.g. mirror neutrinos \cite{flv}, $V_s$ can be non-zero.
We will comment on this case later.) For a weak eigenstate
neutrino ($\nu_{\alpha}, \alpha = e, \mu, \tau$), $V_{\alpha}$ is given
by \cite{nr, ekm2, c}
$$V_{\alpha} = \sqrt{2} G_F N_{\gamma}\left( L^{(\alpha)} -
{A_{\alpha} T^2 \over M_W^2}\right), \eqno (5)$$
where $G_F$ is the Fermi coupling constant, $M_{W}$
is the $W$ boson mass, $A_{\alpha}$ is a numerical factor
given by $A_e \simeq 55$ and $A_{\mu, \tau} \simeq 15.3$ \cite{nr,ekm2}.
The neutrino asymmetry $L^{(\alpha)}$ is given by
$$L^{(\alpha)} = L_{\alpha} + L_{\nu_e} + L_{\nu_{\mu}} +
L_{\nu_{\tau}},
\eqno(6)$$
where $L_{\alpha} = (N_{\alpha} - N_{\bar \alpha})/N_{\gamma}$.
Since we will be interested in the case where the asymmetries are
large (of order $10^{-5}$ or more)
we have neglected the asymmetries in the electrons
and protons/neutrons since these are known to be small ($\sim
10^{-10}$).

For $L^{(\alpha)}$ non negligible, the bounds of Eq.(1) can be
weakened considerably. This is because the oscillation probability
depends on $L^{(\alpha)}$ through the dependence of
$\sin^2 2\theta_m$ on $L^{(\alpha)}$ in Eqs.(4,5).
The condition that the sterile neutrinos not come into
equilibrium is that the interaction rate for
sterile neutrinos is less than the expansion rate,
i.e. $\Gamma_{\nu_s} < H$.
We will assume that there are essentially no sterile neutrinos
initially. The rate of production of
sterile neutrinos is given by
the interaction
rate of ordinary neutrinos multiplied by the probability that
the neutrino collapses to the sterile eigenfunction, i.e.
$$\Gamma_{\nu_s}(t) = \langle P_{\nu_{\alpha}} \rightarrow \nu_s
\rangle_{coll} \Gamma_{\nu_{\alpha}}, \eqno (7)$$
where $\Gamma_{\nu_{\alpha} } = y_{\alpha} G_F^2 T^5$,
($y_e \simeq 4.0$ and $y_{\mu,\tau} \simeq 2.9$) \cite{B}
and $\langle P_{\nu_{\alpha}} \rightarrow \nu_s \rangle_{coll}$
is given by
$$ \langle P_{\nu_{\alpha}} \rightarrow \nu_s \rangle_{coll} =
\sin^2 2\theta_m \langle \sin^2 {x \over L_{osc}^{(m)}}
 \rangle, \eqno(8)$$
where $x$ is the distance between collisions.
Note that $\langle x \rangle \equiv L_{int} = 1/\Gamma_{\nu_{\alpha}}$
where $L_{int}$ is the mean distance between interactions. From
Eq.(4) it is easy to see that the production rate
of sterile neutrinos is significantly suppressed
(because $\sin^2 2\theta_m \ll \sin^2 2\theta_0$)
for temperatures above about 12 MeV (given $\delta m^2
\sim 10^{-2} \ eV^2$),
independently of the magnitude of $L^{(\alpha)}$
(except in a resonance
region where $\sin^2 2\theta_m = 1$,
as we shall discuss later).
Below about 12 MeV, $\sin^2 2\theta_m$ approaches its
vacuum value (unless $L^{(\alpha)}$ is non negligible).
In the standard scenario\cite{B}, it is in this region
where oscillations can occur and potentially bring
the sterile neutrino into equilibrium.

However, for $L^{(\alpha)}$ non negligible, the
production rate of sterile neutrinos can continue to be
suppressed in the region below 12 MeV. Assuming
$|L^{(\alpha)}| > 10^{-5}$ (which we will
discuss later),
the condition $L_{int}/L^{(m)}_{osc} \gg 1$ always holds
off resonance (for temperatures above the decoupling temperature).
This means that
$\langle \sin^2 (L_{int}/L_{osc}^{(m)}) \rangle$ averages
to $1/2$.   Using this result, and Eqs.(7,8) [with
$\sin^2 2\theta_m$ given by Eq.(4)], we can now
calculate the production rate of
sterile neutrinos for the general
case with $L^{(\alpha)}$ non-zero.
Demanding that this interaction rate be less than the expansion
rate, i.e. $\Gamma_{\nu_s}\ \stackrel{<}{\sim} \
H \simeq 5.5 T^2/M_P$,
we find for large $L^{(\alpha)}$ (where the second term in
Eq.(5) can be neglected) and maximal mixing
$$ (\delta m^2)^2\ \stackrel{<}{\sim} \
79 G_F^2 T^2 N_{\gamma}^2
\left[ L^{(\alpha)}\right] ^2
\left({y_{\alpha}M_P G_F^2 T^3 \over 11} - 1\right)^{-1},
\eqno (9)$$
where $M_P \simeq 1.2 \times 10^{22}\ MeV$ is the Planck mass.
Note that in the case where the mixing is not maximal, the bound
is even more stringent.
Clearly, oscillations from ordinary to sterile neutrinos which
occur after the kinetic decoupling temperature ($T_{dec}$)
do  not significantly affect
the energy density of the universe.
For temperatures above $T_{dec}$, the most
stringent bound occurs at
the decoupling temperature. For $\nu_e$, $T_{dec} \simeq 2.6$ MeV
and for $\nu_{\mu, \tau}$, $T_{dec} \simeq 4.4$ MeV, which
leads to the bounds:
$$ |\delta m^2|\ < \ 4 \times 10^2 |L^{(e)}| eV^2, \
\nu = \nu_e$$
$$ |\delta m^2|\ < \ 1.6 \times 10^3 |L^{(\mu, \tau)}| eV^2, \
\nu = \nu_{\mu, \tau}.
\eqno (10)$$
These bounds replace those of Eq.(1) in the case where
$|L^{(\alpha)}|$ is large [$|L^{(\alpha)}| > 10^{-5}$].

In the case of electron neutrino oscillations into sterile
neutrinos a more stringent bound comes by requiring
that the electron neutrinos are not depleted significantly
down to temperatures where the protons and neutrons
go out of equilibrium (which is about $0.8$ MeV in the
standard scenario).
For maximal mixing, the bound is $\delta m^2 < 10^{-8}\ eV^2$ if
$L^{(e)}$ is negligible\cite{B}.
However in the case where $|L^{(e)}|$
is not negligible ($> 10^{-5}$),
the situation is changed. If we demand that
$\sin^2 2\theta_m \stackrel{<}{\sim}\ 1/10$ for $T \ge 0.8$
MeV (which means that the $\nu_e - \nu_s$ oscillations are severely
suppressed for $T \ge 0.8$ MeV)
 we find in the case of maximal mixing that
$$|\delta m^2| \  \stackrel{<}{\sim}
2\sqrt{2} |L^{(e)}| G_F N_{\gamma} T
\  \stackrel{<}{\sim} 3.3 |L^{(e)}|\ eV^2. \eqno (11)$$
This bound for $\nu_e - \nu_s$ oscillations is
clearly more stringent than Eq.(10).

In the above analysis we have assumed that the value of
$L^{(\alpha)}$
is fixed. However in reality
$L^{(\alpha)}$ is in general not constant. Oscillations can
change its value. There are only two regions where
oscillations are important. First, there is the
region near
the high temperature resonance (in this
region $\sin^2 2\theta_m = 1$). The other region where
oscillations can significantly change $L^{(\alpha)}$ is at the
low temperature region $ T \approx T_{dec}$.
Oscillations can change $L^{(\alpha)}$ in this
region because the production rate of sterile neutrinos
is not so strongly suppressed (recall $\sin^2 2\theta_m
\rightarrow \sin^2 2\theta_0$ as $T \rightarrow 0$).
We now consider each of these regions in turn.

We will assume for definiteness that $L^{(\alpha)}$ is
positive (unless explicitly stated otherwise).
The production rate of sterile neutrinos is given by Eqs.(7,8),
with $\sin^2 2\theta_m$ given by Eqs.(4,5).
The condition for $\nu_{\alpha}-\nu_s$ oscillation
resonance ($\theta_m = \pi/4$) is that
$$(V_{\alpha} - V_s) = \Delta_0 \cos 2\theta_0, \eqno (12)$$
implying that the resonance temperature is\cite{c}
$$T_{res}^2 = {L^{(\alpha)} M_W^2 \over A_{\alpha}}
- {\Delta_0 \cos 2\theta_0 M_W^2 \over A_{\alpha}\sqrt{2} G_F N_{\gamma}}.
\eqno (13)
$$
Observe that strictly, the right-hand side of Eq.(13) is a
function of temperature so that
we must solve Eq.(13) for the resonance temperature. However,
it turns out we will only
be interested in the high temperature resonance and
quite large values of $L^{(\alpha)}$, i.e.
$L^{(\alpha)}\ \stackrel{>}{\sim}\ 10^{-5}$,
and in this case the second term on the right-hand side of Eq.(13)
can be neglected (for $\delta m^2 \le 1\ eV^2$).

Observe that for $L^{(\alpha)} > 0,\ \delta m^2 > 0$, there is no
resonance for antineutrinos while for
$L^{(\alpha)} >0,\ \delta m^2 < 0$
the resonance for antineutrinos occurs at very low temperatures.
[Note that for $\delta m^2 \le 10^{-2} eV^2$, $L^{(\alpha)}
\ge 10^{-5}$ this low temperature resonance occurs at
temperatures below the kinetic decoupling temperature and
thus it can be neglected in our analysis.]
Of more importance is the ``high temperature'' resonance
which, for $L^{(\alpha)} > 0$ only occurs for
neutrinos (for $L^{(\alpha)} < 0$ the
high temperature resonance only occurs for antineutrinos).
The effect of the high temperature resonance is rather interesting.
For an initial $L^{(\alpha)}$ less than a certain ``critical''
value (to be determined later)
$L^{(\alpha)} $ will evolve to zero.
This is essentially because the oscillations near the
resonance are so numerous as to continually lower the
value of $|L^{(\alpha)}|$ and hence also
the resonance temperature, so that the system cannot actually
pass through the resonance.

However, for $L^{(\alpha)}$ large enough
there will be a critical point where the expansion of the universe
is more important than the change in $L^{(\alpha)}$ due to oscillations.
This behaviour has been studied numerically in Ref.\cite{ekm2, ekm}.
Below we show how this behaviour can be understood and we derive
an analytic approximation for the critical value of $L^{(\alpha)}$.

If the change in the resonance temperature
due to oscillations is greater than the width of the resonance,
then the resonance will be moved to lower and lower
temperatures, until the lepton number is reduced to
near zero. However if
the change in the resonance temperature due to oscillations
during the resonance
is less than the width of the resonance \cite{f1}, then this will ensure
the system passes briefly through the resonance. Having
passed through the resonance, the value of
$L^{(\alpha)}$ will not change significantly until much
lower temperatures, as discussed earlier.
The condition that the system passes through the
resonance is that
$$\delta T_{res}\ \stackrel{<}{\sim}\ \Delta T, \eqno (14)$$
where $\delta T_{res}$ is the change in the resonance temperature
due to the oscillations as the system passes through the resonance,
while $\Delta T$ is the width of the resonance. The
resonance temperature
$T_{res}$ is related to the lepton number asymmetry
through Eq.(13) so that
$$\delta T_{res} = {M_W \delta L^{(\alpha)} \over
2\sqrt{A_{\alpha} L^{(\alpha)}}}. \eqno (15)$$
Now, $\delta L^{(\alpha)}$ is proportional to
the reaction rate $\Gamma_{\nu_s}$ for
ordinary neutrinos converting into sterile neutrinos multiplied
by the time it takes for the system to pass through the resonance $
\Delta t$, i.e.
$\delta L^{(\alpha)} = -3\Gamma_{\nu_s}\Delta t/4$.
Note that the resonance width measures defined in terms of the
temperature and the time, $\Delta T$ and $\Delta t$, are related to each
other using the time-temperature relation of the early universe:
$t \simeq M_P/11T^2 \ $
implies $\Delta t \simeq - M_P\Delta T/5.5T^3$.
Hence, the condition that the system will pass
through the resonance is that
$${3M_W M_P \Gamma_{\nu_s}\over 44 T^3 \sqrt{A_{\alpha} L^{(\alpha)}}}
\  \stackrel{<}{\sim}  1. \eqno(16)$$
Note that for $L^{(\alpha)} > 10^{-5}$, $T_{res} > 35 $
MeV. It is easy to verify that at the resonance,
$L_{int}/L_{osc}^{(m)} \ll 1$ and hence
 $\sin^2 (L_{int}/L_{osc}^{(m)}) \rangle \simeq
L_{int}^2/L_{osc}^{(m)2}$. Using this result, and Eqs. (7,8),
we can calculate
$\Gamma_{\nu_s}$ at the resonance:
$$\Gamma_{\nu_s}|_{res} = {\Delta^2_0 \over y_{\alpha}G_F^2 T^5}.
\eqno (17)$$
Substituting $\Gamma_{\nu_s}|_{res}$ into Eq.(16), we obtain
$$L_{crit}^{(\alpha)}\ \simeq \left({(\delta m^2)^4 M_P^2 A^9_{\alpha}
\over 3 \times 10^5 y_{\alpha}^2
G_F^4 M_W^{18}}\right)^{1 \over 11}. \eqno (18)$$
Note that $L_{crit}^{(\alpha)}$ is independent of the vacuum
mixing angle $\theta_0$.
For $L^{(\alpha)} < L^{(\alpha)}_{crit}$
the change in the resonance temperature is greater than
the width of the resonance. This means that
the resonance dynamically evolves to later and
later times, with $L^{(\alpha)}$ moving closer
and closer to zero.
For $L^{(\alpha)} > L^{(\alpha)}_{crit}$
the change in the resonance temperature is less
than the width of the resonance, so the system
passes through the resonance.
Having passed through the resonance the value of
$L^{(\alpha)}$ remains approximately unchanged (until
much later times as will be discussed later).
Thus requiring $L^{(\alpha)} > L^{(\alpha)}_{crit}$ we find
$$L^{(e)} \ > \ 9 \times 10^{-6}\
{\rm for}\ \delta m^2 \le 10^{-4} \ eV^2,$$
$$L^{(\mu, \tau)} \ > \ 1.8 \times 10^{-5}\
{\rm for}\ \delta m^2 \le 10^{-2} \ eV^2. \eqno (19)$$

It is useful to compare our analytic expression Eq.(18) with
the numerical work of Ref.\cite{ekm}.
They find numerically that for $\delta m^2 = 10^{-4}\ eV^2$
an initial asymmetry of $L^{(e)}  = 10^{-5}$
remains unchanged on passing through the resonance.
They also obtain
that for $\delta m^2 = 10^{-3}\  eV^2$
an initial asymmetry $L^{(e)}_0 = 10^{-5}$ leads $L^{(e)}$ to evolve
to zero. These results are consistent with our
analytic expression Eq.(18).
[Putting in $\delta m^2 = 10^{-4} \
eV^2$ in Eq.(18), we find $L^{(e)}_{crit}  \simeq 9 \times 10^{-6}
< L^{(e)}_0$
while for $\delta m^2 = 10^{-3} \ eV^2$ we find
$L^{(e)}_{crit} \simeq 2.1 \times 10^{-5} > L^{(e)}_0$.]

As a consistency test,
we should check that the change in $L^{(\alpha)}$
on passing through the resonance is small compared with the initial
value of $L^{(\alpha)}$.
To work out the change in $L^{(\alpha)}$ on passing through the
resonance, we must work out the width of the resonance.
Note that the width of the resonance is larger than
might be expected. Actually, the interaction rate
at the centre of the resonance $\theta_m = \pi/4$ is
equal to the interaction rate anywhere in the region where
$L^{(m)}_{osc} = \Delta_m^{-1} \gg L_{int}$. To see this observe that
the interaction rate is given by,
$$\Gamma_{\nu_s} = \Gamma_{\nu_{\alpha}} \sin^2 2\theta_m
\sin^2\left({L_{int} \over L_{osc}}\right)
 = y_{\alpha} G_F^2 T^5 {\Delta_0^2 \over \Delta_m^2}
\sin^2\left({\Delta_m \over y_{\alpha}G_F^2 T^5}\right).
\eqno (20)$$
Thus, $\Gamma_{\nu_s} = \Gamma_{\nu_s}|_{res}$ [defined in Eq.(17)]
provided $\Delta_m \ll L_{int}^{-1} = y_{\alpha} G_F^2 T^5$.
Using Eqs.(13-15), it is easy to show that
the change in $L^{(\alpha)}$ on passing through the resonance
is related to the resonance width by $\delta L^{(\alpha)}/L^{(\alpha)}
= 2\Delta T/T$.
Calculating $\Delta T$ using $\Delta_m (T_{res} + \Delta T/2)
\simeq y_{\alpha} G_F^2 T^5_{res}$,
we find that $\delta L^{(\alpha)}/L^{(\alpha)}
\le 0.04\ (0.10) $ for
$\nu_{\alpha} = \nu_e$ ($\nu_{\alpha} = \nu_{\mu, \tau}$).
Thus, the the change in $L^{(\alpha)}$ through the resonance
is always at least an order of magnitude smaller than $L^{(\alpha)}$.

Note that the calculation of $L_{crit}^{(\alpha)}$ assumes
that the sterile neutrino has no significant interactions.
In the case of  a neutrino which has only significant
interactions with itself (such as a mirror neutrino \cite{flv}) then
$V_s $ is unequal to zero.
In this case $L^{(\alpha)}$ does not evolve to zero
but evolves so that $V_{\alpha} - V_s \approx 0$. This does
not affect the resulting analysis, since the
interaction rate depends on $V_{\alpha} - V_s$ rather than $V_{\alpha}$.

We now consider the low temperature region $T \approx T_{dec}$
in which oscillations can also potentially erase $L^{(\alpha)}$.
To calculate $\delta L^{(\alpha)}$ in this region, we must integrate
$\delta L^{(\alpha)} = -3(\Gamma_{\nu_s} - \Gamma_{\overline{\nu}_s})
\delta t/4$, using $\Gamma = (1/2)
\Gamma_{\nu_{\alpha}} \sin^2\theta_0/(1 - 2z\cos 2\theta_0  + z^2)$
where $z$ differs between $\nu$ and $\overline{\nu}$ by $L^{(\alpha)}
\to - L^{(\alpha)}$.
Approximating $1 -2z\cos2\theta_0 + z^2$
by $z^2$, which is approximately valid
for $L^{(\alpha)} > 10^{-5}$ and $\delta m^2 \le 10^{-2}\
eV^2\ (10^{-5} \ eV^2$) for
$\alpha = \mu, \tau$ and $T  \ge 4.4 \ MeV$ ($\alpha = e$
and $T \ge 0.8 \ MeV$), we find that
$$ \left( L_{\rm final}^{(\alpha)} \right)^4 \simeq \left( L_{\rm
initial}^{(\alpha)} \right)^4 -
\frac{3\sin^2\theta_0 A_{\alpha} y_{\alpha}
M_P (\delta m^2)^2}{22M_W^2 T_{f}^3}
\ >  \ (10^{-5})^4, \eqno(21)$$
where the last inequality comes by demanding that $L_{\rm final}^{(\alpha)}
> 10^{-5}$ so that the bounds Eqs.(10, 11) remain valid.
Note that the strongest tendency towards
erasure of $L^{(\alpha)}$
occurs at the low temperature end of the integration region.
For $\alpha = \mu, \tau$, $T_f = T_{dec} \simeq 4.4\ MeV$.
For $\alpha = e$, we require $T_f \simeq 0.8\ MeV$ because we
require that $L^{(e)}$ not be erased above the temperature
in which the protons and neutrons are kept in equilibrium.
Evaluating Eq.(21) we find that in the most stringent case of
maximal mixing:
$$L_{\rm initial}^{(e)} > 4 \times 10^{-5}\ {\rm for} \
\delta m^2 \le 10^{-4} \ eV^2,$$
$$L_{\rm initial}^{(\mu, \tau)} > 7 \times 10^{-5}
\ {\rm for} \ \delta m^2 \le 10^{-2} \ eV^2. \eqno (22)$$
These bounds are slightly more stringent (for maximal
mixing) than those
obtained earlier from requiring that $L^{(\alpha)}$ not
be erased due to the high temperature resonance.

Thus, we conclude that if $L^{(\alpha)}$ satisfies
Eqs.(19) and (22)
then $L^{(\alpha)}$ is not erased either at high temperatures
or at low temperatures, and hence for $L^{(\alpha)}$ satisfying
Eqs.(19) and (22) the bounds Eqs.(10, 11) hold. We conclude that
for $L^{(\alpha)}$ satisfying Eqs.(19) and (22)
the sterile neutrino solution to the atmospheric neutrino
anomaly is consistent with BBN. Also, the large angle
MSW $\nu_e - \nu_s$ oscillation solution to the solar neutrino problem
is also consistent with BBN
as is the maximal mixing vacuum
oscillation solution (see e.g. Ref.\cite{flv}).

\vskip 5mm
\centerline{Acknowledgements}
We thank M. Thomson for many enlightening discussions.
This work was supported by the Australian Research Council.


\vskip 1.5cm
\begin{thebibliography}{99}

\bibitem{snd}
GALLEX Collab., Phys. Lett. B327, 377 (1994);
SAGE Collab., Phys Lett. 328 B, 234 (1994).
Kamiokande Collab., Nucl. Phys. B38 (Proc. Suppl.), 55 (1995);
Homestake Collab., {\it ibid}, 47.


\bibitem{ana}
K. Hirata et al, Phys. Lett. B280, 146 (1992);
D. Casper et al, Phys. Rev. Lett. 66, 2561 (1991);
Y. Fukuda, Phys. Lett. 335B, 237 (1994).

\bibitem{lsnd}
LSND Collaboration, Phys. Rev. Lett. 75, 2650 (1995); see
however J. E. Hill {\it ibid.} 75, 2654 (1995).

\bibitem{sn}
See e.g. V. Barger et al., Phys. Rev. Lett. 45, 692 (1980);
A. Yu. Smirnov and J. W. F. Valle, Nucl. Phys. B375, 649
(1992); J. Peltoniemi et al., Phys. Lett. B298, 383 (1993);
D. O. Caldwell and R. N. Mohapatra, Phys. Rev. D48, 3259 (1993);
A. S. Joshipura and J. W. F. Valle, hep-ph/9410259 (1994);
E. Ma and P. Roy, hep-ph/9504342 (1995);
Z. Berezhiani and R. N. Mohapatra, hep-ph/9505385 (1995).

\bibitem{flv}
R. Foot, H. Lew and R. R. Volkas, Phys. Lett. B272, 67 (1991);
R. Foot, H. Lew and R. R. Volkas, Mod. Phys. Lett. A7, 2567 (1992);
R. Foot, Mod. Phys. Lett. A9, 169 (1994);
R. Foot and R. R. Volkas, University of Melbourne Preprint,
UM-P-95/49, hep-ph 9505359, Phys. Rev. D (in press).
The concept of exact parity symmetry was
discussed earlier by T. D. Lee and C. N. Yang,
Phys. Rev. 104, 254 (1956); V. Kobzarev, L. Okun and I. Pomeranchuk,
Sov. J. Nucl. Phys. 3, 837 (1966); S. L. Glashow, Phys. Lett.
B167, 35 (1986); E. D. Carlson and S. L. Glashow, {\it ibid.} B193,
168 (1987).

\bibitem{B}
A. Dolgov, Sov. J. Nucl. Phys. 33, 700 (1981);
R. Barbieri and A. Dolgov, Phys. Lett. B237, 440 (1990);
Nucl. Phys. B349, 743 (1991);
K. Kainulainen, Phys. Lett. B244, 191 (1990);
K. Enqvist, K. Kainulainen and M. Thomson, Nucl. Phys. B373, 498
(1992); J. Cline, Phys. Rev. Lett. 68, 3137 (1992);
X. Shi, D. N. Schramm and B. D. Fields,
Phys. Rev. D48, 2568 (1993).

\bibitem{babu}
There is another way to weaken BBN bounds on active to sterile
oscillations in theories with a Majoron; see
K. S. Babu and I. Z. Rothstein, Phys. Lett. B275, 112 (1992).

\bibitem{c2}
Actually $L_{osc}$
is the oscillation length divided by $2\pi$.

\bibitem{msw}
The propagation of neutrinos in matter was first studied by:
L. Wolfenstein, Phys. Rev. D17, 2369 (1978); D20, 2634 (1979);
S. P. Mikheyev and A. Yu. Smirnov, Nuovo Cim. C9, 17 (1986).


\bibitem{nr}
D. Notzold and G. Raffelt, Nucl. Phys. B307, 924 (1988).

\bibitem{ekm2}
K. Enqvist, K. Kainulainen and J. Maalampi,
Nucl. Phys. B349, 754 (1991); Phys. Lett. B 249,
531 (1990).

\bibitem{c}
For antineutrinos, the equation is the same as below, except that
$L^{(\alpha)} \rightarrow -L^{(\alpha)}$.

\bibitem{ekm}
K. Enqvist, K. Kainulainen and J. Maalampi,
Phys. Lett. B244, 186 (1990).

\bibitem{f1}
It is possible to show that this condition
is equivalent to requiring that
the change in the interaction rate due to oscillations (at
resonance) is
less than the change in the interaction rate due to
the expansion of the universe.
In other words we require that the expansion of
the universe takes the system out of the resonance
region and not the oscillations.

\end{thebibliography}
\end{document}